# Comparing Radiality Constraints for Optimized Distribution System Reconfiguration

Pablo Cortés, Alejandra Tabares, Fredy Franco, Astrid Xiomara Rodríguez , David Álvarez-Martínez

**Abstract**

The reconfiguration of electrical power distribution systems is a crucial optimization problem aimed at minimizing power losses by altering the system's topology through the operation of interconnection switches. This problem, typically modelled as a mixed-integer nonlinear program (MINLP) demands high computational resources for large-scale networks and requires specialized radiality constraints for maintaining the tree-like structure of distribution networks. This paper presents a comprehensive analysis that integrates and compares the computational burden associated with different radiality constraint formulations proposed in the specialized literature for the reconfiguration of distribution systems (RDS). By using consistent hardware and software setups, the performance of these constraints was evaluated across several well-known test cases, including systems with 14, 33, 84, 136, and 417 buses. Our findings reveal significant differences in computational efficiency depending on the chosen set of radiality constraints, providing valuable insights for optimizing reconfiguration strategies in practical distribution networks.

Keywords: Electrical distribution systems, mixed-integer non-linear programming, reconfiguration.

**Nomenclature**

*Variables:*

| | |
|---|---|
| $f_{ij}$ | Amount of commodity flowing through branch $(i,j)$ |
| $f_{ij}^{k}$ | Amount of commodity $k$ flowing through branch $(i,j)$ |
| $I_{ij}$ | Electrical current flowing through branch $(i,j)$ |
| $P_{ij}$ | Active power flowing through branch $(i,j)$ |
| $Q_{ij}$ | Reactive power flowing through branch $(i,j)$ |
| $P_i^D$ | Active power demand at bus $i$ |
| $P_i^S$ | Active power generation at bus $i$ |
| $Q_i^D$ | Reactive power demand at bus $i$ |
| $Q_i^S$ | Reactive power generation at bus $i$ |
| $V_i$ | Voltage level of bus $i$ |
| $y_{ij}$ | Binary variable representing the state of switch of branch $(i,j)$. |
| $\Delta_{i,j}^v$ | Voltage slack through branch $(i,j)$ |

*Parameters*

| | |
|---|---|
| $I_{ij}^{ub}$ | Electrical current upper limit for branch $(i,j)$ |
| $M$ | Big constant number |
| $R_{ij}$ | Electrical resistance of branch $(i,j)$ |
| $X_{ij}$ | Inductive reactance of branch $(i,j)$ |
| $V_i^{ub}$ | Voltage upper limit of bus bar $i$ |
| $V_i^{lb}$ | Voltage lower limit of bus bar $i$ |
| $Z_{ik}$ | Electrical impedance of branch $(i,j)$ |

Sets

| | |
|---|---|
| $N$ | Bus set |
| $N_s$ | Substation set |
| $\Omega$ | Set of branches |
| $N_d$ | Set of buses with active or reactive demand different from zero |
| $N_{d0}$ | Set of buses with active and reactive demand equal to zero |

## 1. Introduction

The complexity of electrical distribution systems is continually evolving, driven by rising costs and increasing energy demand. These changes often lead to a decline in the quality of power supplied to consumers, making efficient operation crucial [1]. In this context, the Electrical Power Distribution System Reconfiguration (RDS) strategy has emerged as an effective method for optimizing system performance and reducing operational costs [2]. RDS is particularly attractive due to its low implementation costs, as it leverages existing network resources to achieve significant technical and economic benefits [2].

RDS refers to the process of modifying the topology of a power distribution network to optimize its operation based on various objectives, the most common of which is active power loss. The network topology is adjusted by opening and closing sectional switches, which can be controlled either manually or remotely, aiming for a radial topology of the distribution network [3]. A radial topology refers to a configuration in which the distribution network resembles a tree, with each bus in the network being supplied by a single preceding bus. This configuration facilitates fault isolation and maintenance activities as well as efficient operation. Although electrical distribution systems are typically constructed with a meshed topology, the distribution system operator must be able to achieve a radial topology that minimizes power losses.

RDS is inherently a combinatorial optimization problem, where the state of the switches is modelled as a binary control variable. The complexity of this problem scales with the size of the distribution system, typically characterized by the number of switches within the network. The literature offers various strategies to solve the RDS problem, with approaches ranging from mathematical programming to artificial intelligence; for a comprehensive review on the different techniques used to solve EPSDR, see reference [3].

Mathematical programming approaches offer a straightforward solution for non-linear optimization problems, including continuous and integer programs. However, their weakness lies in their inefficiency in solving extensive combinatorial search problems, making them computationally expensive and, in some cases, inapplicable to a real-world

RDS due to unaffordable computational requirements [4]. Within mathematical programming techniques, one of the major challenges has been the definition of radiality conditions for the network. The first attempt to enforce radial conditions for the EPSDR was made by Merlin et al [5] in 1974; in this pioneering work, radial conditions were ensured by observing that, in a radial topology, the total number of active circuit branches must equal the number of buses minus one. However, the oversimplification of the electrical constraints, assuming pure resistive circuit branches and direct current, led to ongoing research in the subsequent decades.

Twenty years later, a linear program model was developed by [6], formulating the RDS as a minimum-cost flow optimization problem; in this mathematical program, voltage constraints on the network were disregarded, although current limits on each circuit branch were included. The radial topology of the solution was ensured by forcing the number of active circuit branches to equal the number of buses minus one. A decade afterwards, the RDS problem was formulated by [7] as a mixed-integer linear optimization problem by approximating the quadratic power flows through circuit branches with piecewise linear functions. Radiality conditions were ensured by defining paths; a path is a set of circuit branches that connect a specific bus to the substation. In this way, the radial topology was achieved if a particular bus has a single active path. This mixed-integer linear optimization program allowed for finding the optimal configuration of the 32 and 69-bus test systems, although power losses were underestimated due to the linearization of power flows.

An improvement in the radial topology conditions was achieved in 2012 by Jabr et al. [8]. This work developed a mixed-integer conic programming (MICP) formulation for RDS. To ensure the radial topology of the network, the study introduced the so-called spanning tree constraints for the first time. These constraints are based on the observation that every bus in the distribution network, except for the substation, must have exactly one parent (i.e., a feeder bus). This condition was represented by introducing two additional binary variables for each circuit branch. The proposed model was tested on a distribution network with 830 buses, and it was found that the MICP formulation can find a feasible solution more quickly compared to a mixed-integer linear programming (MILP) formulation. However, no study has been conducted to elucidate the advantages of using spanning tree constraints versus other radiality conditions in terms of computational resources.

To characterize radial topology constraints, in parallel with [8], [9], a generalization of the radiality conditions developed by Merlin was introduced by Lavorato et al. [10]. This work by Lavorato et al. performed a critical analysis and the proposal for incorporating the radiality constraints in mathematical models. To address transfer buses (buses with zero generation or demand), an additional binary variable was introduced to represent the utilization of the transfer buses. Seeking alternative radiality conditions that would work for the general case where the network has transfer buses, Jabr [11] formulated in 2013 the so-called Single Commodity Flow constraints. This set of constraints are based on assuming unitary fictitious demands of a certain commodity at each of the network buses that could be supplied by the substations. The so-called multi-commodity flow constraints is an extension of this technique [11], which uses fictitious demand and supplies of different commodities for each bus. However, the main weakness of this work was the assessment carried out only for small and medium test systems (23, 54, and 136 buses).

An improvement to the radiality conditions generalized by Lavorato [10] was introduced by Franco et al. [9] in 2013. They demonstrated that using two binary variables to represent the state of a switch, instead of a single variable, significantly enhanced the computational efficiency of the off-the-shelf software (CPLEX) used to solve the optimization problem. Nonetheless, this new formulation of the radiality conditions was not compared to either the spanning tree constraints or the commodity flow constraints.

By combining approaches used in [11], a novel set of radiality constraints was created by mixing single commodity flow with spanning tree constraints [12] in 2020; it was tested on distribution networks comprising 32, 83, 135, and 201 buses. The computation time was compared between this new set of constraints and the traditional single commodity flow constraints. The study concluded that the proposed set of constraints was more efficient than the single commodity flow constraints previously proposed by Jabr [11]. Nevertheless, the study did not evaluate the efficiency of the proposed constraints against multi-commodity flow constraints or standalone spanning tree constraints.

Four years later, a simplified form of the spanning tree constraints was proposed by [13] utilizing only two binary variables instead of the three used in the original formulation by Jabr. This new formulation was compared to the radiality constraints proposed by Lavorato [10] and Franco [9] using test systems with 14, 33, 84, 136, and 416 buses. The study concluded that the spanning tree constraints were more efficient than the other two formulations. Nonetheless, the study did not include a comparison against single or multi-commodity flow constraints.

Although numerous radiality formulations for distribution systems have been proposed in the literature, no comprehensive comparative study has been conducted to evaluate their respective computational burdens. This paper fills that gap by analyzing experimentally the performance of several radiality constraint sets, aiming to reduce computational time when using commercial solvers. The formulations were implemented in Gurobi and tested on standard benchmark systems with 14, 33, 136, and 417 buses, widely used in the scientific community.

This paper is organized as follows: Section 2 defines the mathematical modeling of the distribution system reconfiguration problem, and the sets of auxiliary constraints applied to the implemented models; Section 3 presents the methodology used and the results for the test systems; and Section 4 provides the conclusions of this work.

## 2. Mathematical Model for the RDS Problem

In this section, the optimization model originally proposed by [14] for RDS is presented. Subsequently, the different formulations of radiality conditions considered in this study are introduced, along with some proposed improvements, which will also be evaluated in the results section.

### 2.1 Optimization Model for RDS

In a basic and generalized way, the optimization model for the RDS problem is formulated as a mixed-integer nonlinear programming (MINLP) problem, as specified in [9]. The model is designed to minimize the electrical losses in the system due to the Joule effect, subject to a set of constraints that ensure the physical and operational integrity of the distribution network. The formulation is presented through equations (1)–(10).

The objective function, given by equation (1), aims to minimize the total power losses across all branches of the distribution network. These losses are calculated as the sum of the products of the resistance $R_{k,i}$ and the square of the current $I_{k,i}$ flowing through each branch $(k, i)$ in the network. Equations (2) and (3) ensure the balance of active and reactive power at each bus $i$ within the network, holding Kirchhoff's current law,. Here, $P_i^S$ and $Q_i^S$ represent the active and reactive power generation at bus i, respectively, while $P_i^D$ and $Q_i^D$ denote the active and reactive power demands at bus i. The terms $R_{i,j}$ and $X_{i,j}$ are the resistance and inductive reactance of the branch between buses $i$ and $j$, respectively.

$$\min \sum_{(i,j)\in\Omega} R_{k,i} \cdot I_{i,j}^{sqr} \quad (1)$$

$$\sum_{(k,i)\in\Omega} P_{k,i} + P_i^S = P_i^D + \sum_{(i,j)\in\Omega} (P_{i,j} + R_{i,j} \cdot I_{i,j}^{sqr}) \;\forall\; i \in N \quad (2)$$

$$\sum_{(k,i)\in\Omega} Q_{k,i} + Q_i^S = Q_i^D + \sum_{(i,j)\in\Omega} (Q_{i,j} + X_{i,j} \cdot I_{i,j}^{sqr}) \;\forall\; i \in N \quad (3)$$

$$V_i^{sqr} - 2(R_{i,j}P_{i,j} + X_{i,j}Q_{i,j}) - Z_{i,j}^2 I_{i,j}^{sqr} = V_i^{sqr} + \Delta_{i,j}^v \;\forall\; (i,j) \in \Omega \quad (4)$$

$$\left|\Delta_{i,j}^v\right| \leq b^v \cdot (1 - y_{i,j}) \;\forall\; (i,j) \in \Omega \quad (5)$$

$$V_j^{sqr} I_{i,j}^{sqr} \geq P_{i,j}^2 + Q_{i,j}^2 \;\forall\; (i,j) \in \Omega \quad (6)$$

Equation (4) models the voltage drops between two buses $i$ and $k$ connected by a conductor, incorporating both active and reactive power transport and the associated losses, in which $V_i$ and $V_k$ are the voltage levels at buses $i$ and $k$, and $Z_{k,i}$ is the impedance of the branch $(k,i)$, respectively. The term $\Delta_{k,i}^v$ represents the voltage slack through the branch.

Constraint (5) imposes a constraint on the voltage difference between two buses not connected by a circuit branch, allowing it to vary based on the state of the binary variable $y_{k,i}$. Here, $\Delta V_{k,i}^{ub}$ and $\Delta V_{k,i}^{lb}$ are the upper and lower bounds for the voltage slack across the branch. Finally, constraint (6) ensures that the current flowing through a branch is consistent with the corresponding active and reactive power flows, considering the voltage at the receiving bus.

$$V_i^{lb} \leq V_i \leq V_i^{ub} \;\forall\; i \in N \quad (7)$$

$$I_{k,i}^{lb} \leq I_{k,i} \leq I_{k,i}^{ub} \cdot y_{ki} \;\forall\; (k,i) \in \Omega \quad (8)$$

Constraint (7) sets technical limits on the voltage values at the different buses within the distribution system. Constraint restricts the current flow to active circuit branches, ensuring that current can only flow through a branch if its associated switch is closed (i.e., active). Where $I_{k,i}^{lb}$ and $I_{k,i}^{ub}$ represent the lower and upper bounds on the current through branch $(k, i)$, respectively, and $y_{k,i}$ is a binary variable representing the state of the switch (1 if closed, 0 if open).

$$\sum_{(i,j)\epsilon\Omega} y_{ij} \leq |N| - |N_s| \tag{9}$$

$$y_{ij} \in \{0,1\} \ \forall \ (i,j) \in \Omega \tag{10}$$

Traditionally, equations (9) and (10) enforce the radiality of the network joining with equations (2) and (3). Equation (9) limits the number of active branches to ensure that the network remains a tree structure. Wherein $|N|$ is the total number of buses, and $|N_s|$ is the number of substations in the network. Together, these constraints ensure that the RDS problem is solved within the framework of a radial distribution network, minimizing power losses while maintaining operational.

*2.2 Radiality*

The concept of **radiality** in power distribution systems is a critical topic that has been extensively studied due to its direct impact on the operational efficiency and safety of electrical networks. In a radial network, the power distribution system has a tree-like structure where each bus is connected in such a way that there is only one path between any two buses, ensuring a simple and loop-free topology. Radiality is essential because it simplifies the control, protection, and fault isolation processes, making it easier to manage the network, especially during maintenance or in the event of a fault.[15].

The study of radiality is crucial because it directly influences the reliability and efficiency of RDS, which is a key optimization problem for minimizing power losses, balancing loads, improving voltage profiles, and maintaining service continuity. Power distribution networks are traditionally designed as meshed systems but are operated in radial configurations to minimize fault currents and facilitate the use of overcurrent protection schemes. Radiality ensures that protective devices such as circuit breakers and relays can function properly to isolate faults, protecting both the equipment and the consumers.

Moreover, radiality is a fundamental aspect in most optimization problems related to distribution power systems. Whether the objective is optimal power flow, network restoration, fault management, or distributed generation integration, radiality constraints must be enforced to ensure that the distribution system operates in a stable and efficient manner. Without radiality, the network could develop loops that complicate fault isolation, cause voltage instabilities, or overload protection equipment [12].

The reconfiguration problem is a core challenge in power distribution system optimization, where the goal is to alter the topology of the network by opening or closing sectionalizing switches to achieve a specific objective, typically minimizing power losses or improving load balancing. The reconfiguration problem becomes particularly important because it is one of the most effective low-cost strategies available to operators for enhancing the technical performance of the distribution network. Since reconfiguration can be done without significant new investments, it is an attractive solution for improving network efficiency using existing infrastructure.

In graph theory, radiality implies that the network is a tree, meaning it is a connected graph with no cycles. A network with $|N|$ buses must have $||N| - 1$ active edges (circuit branches) to maintain a tree structure. This condition prevents the formation of loops and ensures that every bus (except the substation) is connected to exactly one other bus, ensuring that there is only one path between any two buses. Mathematically, this can be enforced through the following condition:

$$\sum_{(i,j)\epsilon\Omega} y_{ij} \leq |N| - 1 \qquad (11)$$

where $y_{ij}$ is a binary variable representing the status of the switch between buses i and j (1 if the switch is closed, allowing power flow, and 0 if it is open).

The complexity of maintaining radiality while solving the reconfiguration problem lies in its combinatorial nature. For a distribution network with $\rho$ sectional switches, there are $2^\rho$ possible configurations. Although most of these configurations do not form valid radial networks, as show in [10], searching for the optimal configuration becomes computationally expensive, especially for large networks.

One of the primary challenges is that there are necessary mathematical conditions for ensuring radiality (such as the number of active circuit branches being equal to the number of buses minus one), but these conditions are not always sufficient. For example, even if the number of active switches matches the number of buses minus one, the resulting network could still contain loops or disconnected components. Therefore, more sophisticated mathematical formulations are required to guarantee radiality and ensure that the solution respects the physical and operational constraints of the system.

Traditionally, the distribution network has been modelled as either a directed [9] or undirected graph [12]; a tree derived from this graph is obtained by opening or closing the sectional switches available in the circuit branches, ensuring that each bus is connected to only one parent bus. The RDS aims to minimize active power losses in the distribution system, subject to Kirchhoff's voltage and current laws, as well as voltage and current limits for buses and circuit branches, respectively, and radiality constraints [10]. The decision variables controlled by the distribution system operator are the binary states {0,1} of the sectional switches.

The size of the optimization problem is directly related to the number of sectional switches installed in the distribution network; for a distribution network with $\rho$ sectional switches, the total number of possible configurations will be $2^\rho$ [9]. Most of these configurations will generate infeasible solutions, either because they do not form trees or because they violate electrical constraints. As observed, the presence of binary variables modeling the state of the sectional switches renders the RDS a combinatorial problem

[16]. Additionally, the nonlinear nature of the electrical network representation further complicates the solution process. Consequently, mathematical programming approaches often face prohibitive computational times when addressing large distribution systems. Therefore, finding efficient formulations is of paramount importance for the electrical industry. In the following several radiality formulations are described in detail.

### 2.2.1 Parent-Child Relations

The Parent-Child relations are a set of complementary constraints [12], [9], given by inequality (12); this constraint can also be understood in the sense that only current and power can flow in a single direction along each circuit branch of the distribution network [9], as is schematically shown in Figure 1. The complete mathematical program using this constraint is composed by objective function (1) and constraints (2) -(10) and (12).

$$y_{ij} + y_{ji} \leq 1 \; \forall \; (i.j) \in \Omega \tag{12}$$

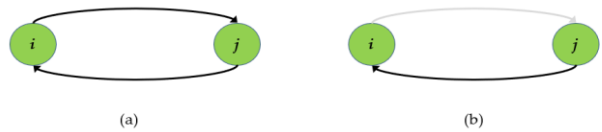

*Figure 1. Effect of Parent child relationships; current can flow in only one direction, as shown in (b)*

### 2.2.2 Spanning Tree Constraints

The constraints of the minimum spanning tree can be formulated according to (13), (14) y (15). Constraint (13) ensures that for all buses with non-zero demand (set $N_d$), there is a connection to the substation, either directly or through other buses. Constraint (14) allows buses with zero demand (set $N_{d0}$) to be connected or not connected to the substation, either directly or through other buses. Finally, constraint (15) ensures that the substation does not receive power from any other buses. It is worth noting that the group of constraints (13), (14), (15) can replace constraints (9) and (12); thus, the complete mathematical program is composed by objective function (1) and constraints (2)-(8), (10), and (13) - (15). Figure 5 shows the effect of spanning tree constraints for a bus with power demand different from zero. It is important to highlight that spanning three constraints guarantee radiality together with KVL and KCL (constraints (2) - (4)), by ensuring that the necessary amount of active and reactive power is delivered to each demand node.

### 2.2.3  Single Commodity Flow Constraints

To construct this set of complementary radiality constraints, it is assumed that at each non-substation bus with non-zero demand, there is a fictitious demand of 1 for a certain commodity; likewise, a fictitious flow $f_{ij}$ (of continuous nature) is assumed on each of the arcs composing the network, allowing this flow to be active only if the breaker of the corresponding circuit branch is active [12].

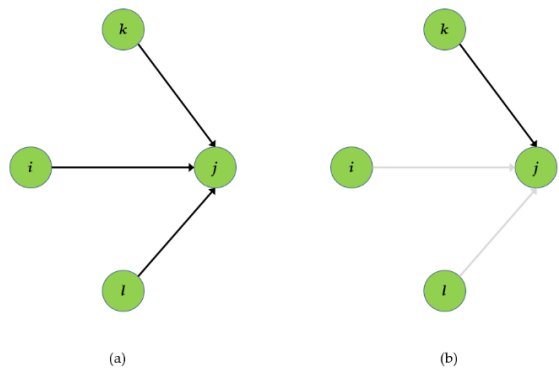

Figure 2. Effect of spanning tree constraint; bus j can only be fed by one bus (in this case k bus)

$$\sum_{i|(i,j)\in\Omega} y_{ij} = 1 \quad \forall\, j \in N_d \tag{13}$$

$$\sum_{i|(i,j)\in\Omega} y_{ij} \leq 1 \quad \forall\, j \in N_{d0} \tag{14}$$

$$\sum_{i|(i,j)\in\Omega} y_{ij} = 0 \quad \forall\, j \in N_s \tag{15}$$

The flow balance equations then allow writing the set of constraints (16), (17), and (18); constraints (16) and (17) correspond to the fictitious flow balance for a bus with demand of the commodity of 1 and 0 respectively; constraint (18) allows the fictitious flows to be active only on the circuit branches activated by the switches of the distribution network.

$$\sum_{j|(i,j)\in\Omega} f_{ij} + 1 = \sum_{k|(k,i)\in\Omega} f_{ki} \quad \forall\, i \in N_d \tag{16}$$

$$\sum_{j|(i,j)\in\Omega} f_{ij} = \sum_{k|(k,i)\in\Omega} f_{ki} \quad \forall\, i \in N_{d0} \tag{17}$$

$$f_{ij} \leq M y_{ij} \quad \forall\, (i,j) \in \Omega \tag{18}$$

Figure 3 shows schematically the main principle behind single commodity flow constraints for a network with six demand buses. A supply of six is assigned to the substation, while a demand of 1 is assigned to each other bus.

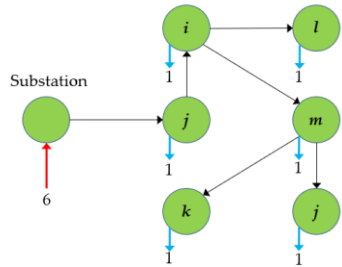

*Figure 3. Main principle of SCF constraints. A supply of six is assigned to the substation, equal to the number of nodes. A demand of 1 is assigned to each other node.*

### 2.2.4 Single Commodity Flow + Spanning Tree Constraints

By combining the constraints of the Single Commodity Flow with those of the minimum spanning tree it is generated the group of complementary constraints called SCF+ST [12]; this group is composed by constraints (13)–(18) [12].

### 2.2.5 Multi-Commodity Flow Constraints

In the case of Multi-Commodity Flow, a fictitious demand of 1 for commodity $k$ is assumed at bus $k$ if the bus has non-zero power demand, and a demand of 0 for commodity $k$ at the rest of the non-substation buses [17].

$$\sum_{j|(i,j)\in\Omega} f_{ji}^k - \sum_{j|(i,j)\in\Omega} f_{ij}^k = -1 \ \forall \ k \in N_d, i \in N_s \tag{19}$$

$$\sum_{j|(j,k)\in\Omega} f_{jk}^k - \sum_{j|(j,k)\in\Omega} f_{kj}^k = 1 \ \forall \ k \in N_d \tag{20}$$

$$\sum_{j|(i,j)\in\Omega} f_{ji}^k - \sum_{j|(i,j)\in\Omega} f_{ij}^k = 0 \ \forall \ k \in N_d, i \in N\setminus\{N_s, k\} \tag{21}$$

$$0 \leq f_{ij}^k \leq y_{ij}, 0 \leq f_{ji}^k \leq y_{ji}, \forall \ k \in N\setminus N_s, (i,j) \in \Omega \tag{22}$$

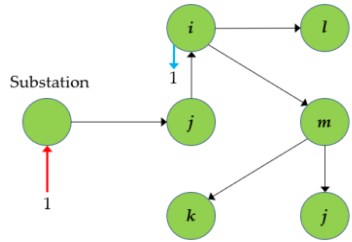

*Figure 4. Main principle behind MCF constraints. A supply of 1 of commodity i is assigned to the substation, while a demand of 1 of the same commodity is assigned to bus i.*

Constraints (19)-(22) form the group of Multi-Commodity Flow constraints; In particular, constraint (19) ensures that one unit of commodity $k$ flows from the substation; constraint

(20) ensures that one unit of commodity k is delivered at bus $k$. Constraint (21) allows the use of all buses other than the substation and bus $k$ as pass-through buses for commodity $k$, while constraint (22) allows the flow of commodity k only through those circuit branches that are switched. Figure 4 shows the main principle behind MCF constraints; a supply of 1 for commodity $i$ is assigned to the substation, while a demand of 1 is assigned to bus $i$. This process is repeated for each bus, using different commodities.

## 3. Enhancing Radiality Constraints for Computational Efficiency

The challenge with enforcing radiality in large distribution networks lies in balancing the precision of the model with computational complexity. To address this, researchers have proposed hybrid approaches that combine the strengths of different auxiliary constraint formulations. Two such combined formulations are discussed below.

### 3.1 Multi-commodity flow plus spanning tree

Combining the constraints of Multi-Commodity Flow with those of the minimum spanning tree generates the group of complementary constraints called MCF+ST, (13)–(15), (19)–(22). This hybrid approach leverages the flow conservation properties of MCF while utilizing the structural advantages of ST constraints. By combining these two methods, the MCF+ST formulation achieves better computational performance in medium-to-large distribution systems, as it reduces the number of possible configurations that need to be evaluated during optimization. This is particularly beneficial in networks with more than 100 buses, where the complexity of the SCF model alone may be prohibitive.

In the MCF+ST formulation, the flow of fictitious commodities ensures that power is delivered radially, while the spanning tree constraints guarantee that the network remains radial. This combination has been shown to reduce computational time without sacrificing solution quality.

*3.2 Multi-commodity flow plus Single commodity flow*

Combining the constraints for Single Commodity Flow and Multi-Commodity Flow yields the group of complementary constraints (16)–(22). Combining the Single Commodity Flow (SCF) and Multi-Commodity Flow (MCF) introduces a powerful hybrid approach that takes advantage of both techniques to ensure radiality in distribution networks. The combination of SCF and MCF allows for more robust control over power flows in the network by leveraging the simplicity of SCF's flow conservation principles and the detailed flow management of MCF across different commodities.

In the SCF formulation, the flow of a single fictitious commodity is assumed across the entire network, ensuring that each bus with demand receives exactly one unit of this commodity, thus maintaining a radial structure. On the other hand, MCF assigns a unique fictitious commodity to each bus, providing more detailed control and ensuring that each demand bus is supplied by exactly one path from the substation.

## 4. Test and Results

In this section, the methodology used to compare the groups of complementary constraints described in Section 2 is described along with the test instances used for this purpose; a total of five instances, with 14, 33, 84, 133, and 417 buses are employed to test the different radiality formulations stablished in the previous section. Subsequently, the findings are presented and discussed.

*4.1 Methodology*

Figure 5 schematically illustrates the methodology employed in the present study. The datasets for the used instances are available in [18]. These datasets encompass information regarding both active and reactive power demands, as well as voltage limits for each bus comprising the distribution network, alongside details pertaining to circuit branches, specifically encompassing resistance and inductive reactance values, and current limits. These datasets are processed utilizing the Python open-source library Networkx to generate a graph-based data structure. Subsequently, this structure is leveraged to construct optimization models within an off-the-shelf optimizer (**Gurobi 11**). The base model, as delineated in Section 2, comprises constraints inherent to load flow in distribution systems, alongside the radiality condition proposed by [5]. The group of models termed "auxiliary constraints" alludes to modifications of the base model with complementary radiality constraints, as expounded upon in detail in Section 2.

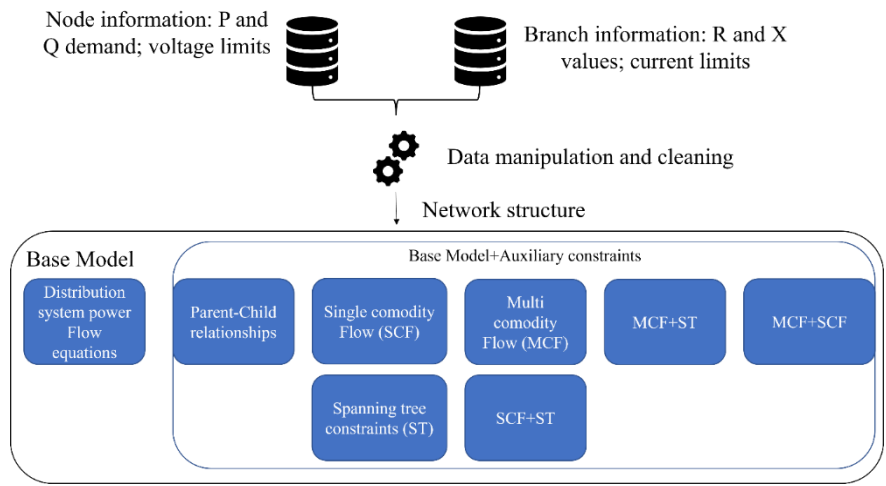

*Figure 5. Methodology's scheme*

### 4.2 Test Systems

The assessment of the different radiality representations stated in section 2 was conducted through examination of five distinct distribution systems retrieved from [18]. Each system varies in terms of the quantity of branches and accompanying details regarding the loads at individual bus bars and resistance and reactance of circuit branches. Table 1 gives comprehensive information pertaining to the number of switches, substation buses, and base voltages and powers for each system under consideration. Conventionally, literature engages with datasets of varying sizes: small distribution systems (comprising 13 and 33 buses), medium distribution systems (with 84, 119, and 136 buses), and large distribution systems (with 417 buses). To ensure comparability with recent literature findings, careful selection of available distribution systems was undertaken, given their frequent utilization in this research domain [19], [9]. These instances serve as standard references for evaluating and contrasting the efficacy of different optimization algorithms and approaches to RDS.

*Table 1. Distribution systems considered for computational study.*

| Number of buses | Number of switches | Reference bus | Base Voltage (kV) | Base apparent power (MVA) |
|---|---|---|---|---|
| 14 | 16 | 14 | 23.0 | 100 |
| 33 | 37 | 1 | 12.66 | 10 |
| 84 | 96 | 84 | 11.4 | 10 |
| 136 | 156 | 1 | 13.8 | 100 |
| 417 | 473 | 1 | 10.0 | 10 |

*4.3 Discussion of results*

In Table 2, the computational time results are presented for each of the five cases considered in this study, as well as for the different groups of auxiliary constraints studied. The various study cases were solved using a virtual machine Intel® Xeon® Gold 6230 R CPU @2.1 GHz with 32 processors and 128 GB of RAM running Ubuntu Linux operating system. The computational time limit was set at 84 hours. As shown in Figure 5, the base model, formed by constraints(2)– (10), fails to found a feasible solution solve the RDS for the instances with 133 and 417 buses due to the presence of buses with zero active and reactive power demand. In contrast, the sets of auxiliary constraints ST, SCF, MCF, SCF+MCF, SCF+ST, and MCF+ST can successfully solve these instances, as they do not rely solely on constraint (9) to guarantee a radial solution.

However, specifically for the 417-bus instance, none of the groups of auxiliary constraints achieve optimality within the fixed time limit. Furthermore, for the instances with 14, 33, 84, and 136 buses, the ST auxiliary constraints demonstrate the best performance in terms of computational time.

*Table 2. Comparative results*

| Model | Variable | \multicolumn{5}{c}{Test Systems} | | | | |
|---|---|---|---|---|---|---|
| | | 14 | 33 | 84 | 136 | 417 |
| Base model | Time (s) | 0,12 | 1,75 | 9.85 | * | * |
| | Power Losses (kW) | 605.92 | 139.55 | 469.87 | * | * |
| P-C | Time (s) | 0,14 | 0,53 | 7,68 | * | * |
| | Power Losses (kW) | 605.92 | 139.55 | 469.87 | * | * |
| SCF | Time (s) | 0,17 | 1,04 | 3,59 | 403,43 | 302400 |
| | Power Losses (kW) | 605.92 | 139.55 | 469.87 | 280.14 | 585. 35 |
| | GAP | 0% | 0% | 0% | 0% | 9,08% |
| SCF+ST | Time(s) | 0,11 | 0,7 | 1,91 | 44,69 | 302400 |
| | Power Losses (kW) | 605.92 | 139.55 | 469.87 | 280.14 | 583.12 |
| | GAP | 0 | 0 | 0 | 0 | 7,99% |
| MCF | Time (s) | 0,16 | 1,71 | 9,47 | 329,44 | 302400 |
| | Power Losses (kW) | 605.92 | 139.55 | 469.87 | 280.14 | 582.55 |
| | GAP | 0% | 0% | 0% | 0% | 15% |
| MCF+ST | Time (s) | 0,17 | 1,34 | 6,50 | 188,28 | 302400 |
| | Power Losses (kW) | 605.92 | 139.55 | 469.87 | 280.14 | 581,75 |
| | GAP | 0% | 0% | 0% | 0% | 2,68% |
| MCF+SCF | Time (s) | 0,22 | 1,77 | 12,07 | 36806 | 302400 |
| | Power Losses (kW) | 605.92 | 139.55 | 469.87 | 280.14 | 582.89 |
| | GAP | 0% | 0% | 0% | 0% | 17,5% |
| ST | Time (s) | 0,09 | 0,38 | 1,34 | 20,51 | 302400 |
| | Power Losses (kW) | 605.92 | 139.55 | 469.87 | 280.14 | 582.10 |
| | GAP | 0% | 0% | 0% | 0% | 4,72% |

GAP: Solver gap (%)   P-C: Parent-Child relationship   * No feasible solution found

*Table 3. Best known solutions for test systems*

| Case | Best known solution (kW) | Reference |
|---|---|---|
| 14 | 605.9 | [13] |
| 33 | 139.55 | [9] |
| 83 | 469.87 | [19] |
| 133 | 280.14 | [9] |
| 417 | 581.57 | [9] |

Moreover, combining these constraints with SCF or MCF improves computation time compared to using SCF or MCF alone. However, for the 417-bus instance, the group of constraints that performs best in terms of computation time is MCF+ST, surpassing even the exclusive use of the ST group, achieving a gap of 2.68%. On the other hand, specifically for the 14-bus instance, the base model shows better performance in terms of computational time than the SCF and MCF constraint groups.

Table 4. Maximum RAM used in branch and bound tree exploration (MB)

| Model | Test Systems | | | | |
|---|---|---|---|---|---|
| | 14 | 33 | 84 | 136 | 417 |
| Base model | 230.75 | 225.77 | 490.33 | - | - |
| P-C | 166.81 | 217.19 | 366.66 | - | - |
| ST | 167.57 | 207.73 | 236.25 | 665.21 | 121000000 |
| SCF | 165.35 | 249.69 | 206.84 | 1197.86 | 30171.67 |
| MCF | 169.05 | 299.97 | 381.19 | 2947.90 | 22658.93 |
| MCF+ST | 155.49 | 260.96 | 299.07 | 1941.21 | 20891.28 |
| SCF+ST | 164.74 | 209.60 | 390.82 | 547.49 | 30489.79 |
| MSCF+SCF | 171.03 | 263.62 | 413.79 | 1978.42 | 27508.00 |

This behavior may be attributed to the fact that in small instances, adding variables related to fictitious flows and constraints specific to each auxiliary constraint group does not benefit as much as in larger instances, such as the 417-bus case. It is also observed that the SCF+MCF constraint group exhibits the worst performance, with longer computation times even compared to the base model for instances where it can obtain a valid solution (instances of 14, 33, and 84 buses). Therefore, it can be concluded that, for the set of instances analyzed, the SCF+MCF constraint group does not provide any improvement in terms of computational efficiency. The P-C constraint group shows better computational times when compared to the base model but suffers from the same drawback of not guaranteeing the formation of a radial topology when encountering buses with zero demand.

Table 3 displays results reported in the scientific literature for instances of 14, 33, 84, 133, and 417 buses. As observed, the solutions obtained with the different groups of auxiliary constraints considered in this study closely approximate the reported values, confirming the previously discussed results. Furthermore, Moreover, combining these constraints with SCF or MCF improves computation time compared to using SCF or MCF alone. However, for the 417-bus instance, the group of constraints that performs best in terms of computation time is MCF+ST, surpassing even the exclusive use of the ST group, achieving a gap of 2.68%. On the other hand, specifically for the 14-bus instance, the base model shows better performance in terms of computational time than the SCF and MCF constraint groups.

Table 4 presents the maximum RAM consumption by the off-the-shelf software during the exploration process of the branch-and-bound tree. As shown in Figure 6, for instances with 14, 33, and 84 buses, the minimum RAM consumption is achieved by the ST constraint set. However, for larger distribution systems, such as the 136-bus and 417-bus test systems, the ST constraint set exhibits the worst performance in terms of RAM usage. For the 136-bus test system, the SCF+ST constraint set achieves lower consumption, whereas for the 417-bus test system (see Figure 6), the MCF+ST constraint set results in both lower RAM consumption and a reduced optimality gap (see Table 2). This phenomenon may be attributed to the fact that, with a greater number of constraints, it more closely approximates the convex hull of the feasible space, thus reducing the number of nodes generated during the branch and bound algorithm execution.

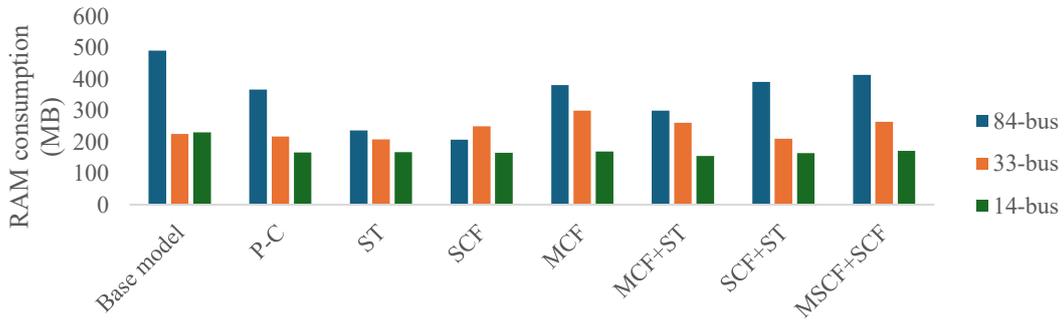

*Figure 6. Maximum RAM used in branch and bound tree exploration (MB) for 14, 33 and 84 bus test systems.*

## 5 Conclusions

The effect of including different groups of auxiliary constraints on the solution of the Electrical Power Distribution System Reconfiguration (RDS) was studied, considering five distribution networks of various sizes that accurately represent the systems commonly used in the literature. Among the analysed constraint groups and the considered networks, the spanning tree constraints (ST) show the best reduction in computational time for instances with 14, 33, 44, and 136 buses.

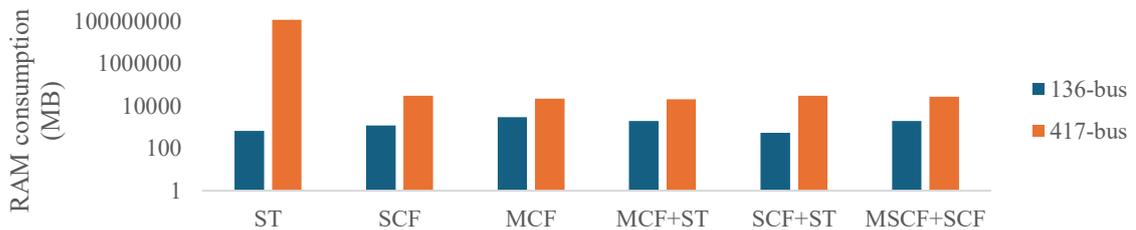

*Figure 7. Maximum RAM used in branch and bound tree exploration (MB) for 136 and 417 bus test systems.*

Only for the 417-bus instance, the auxiliary constraint group Multicommodity Flow together with spanning tree (MCF+ST) exhibits better performance than spanning tree (ST) alone. This can be attributed to the fact that, for such a large system and thus a larger search space, it becomes beneficial to include the entire set of variables related to fictitious flows and their corresponding constraints. While it is not possible to generalize for any distribution network or load profile, it can be stated that the use of ST constraints significantly improves the solution time of RDS. Furthermore, it can be affirmed that for relatively large instances, the use of ST constraints in conjunction with MCF can lead to even lower computation times. Furthermore, the RAM consumption for each constraint group was analysed. The results indicate that for small distribution systems (14, 33, and 84-bus test systems), the ST constraints achieve the lowest RAM consumption. However, for larger distribution systems (133-bus and 417-bus test systems), combinations of formulations such as SCF+ST and MCF+ST outperform the ST constraints in terms of RAM requirements. These results may be significant for even larger distribution systems and real-world applications of RDS, where RAM usage could become a critical limitation.

**Author contribution statement:** Conceptualization, A.T. and F.F.; methodology, A.T.; software, P.C.; validation, A.T., F.F.; formal analysis, A.T and F.F; investigation, P.C.; resources, A.T.; data curation, P.C.; writing—original draft preparation, P.C.; writing—review and editing, A.T and F.F.; visualization, P.C.; supervision, A.T.; project administration, A.T.; funding acquisition, A.T. All authors have read and agreed to the published version of the manuscript.

**Disclosure of interests**: The authors declare they do not have interests to declare

**Funding**: No funding was received

**Data Availability Statement**: The data that support the findings of this study are openly available inhttps://www.feis.unesp.br/#!/departamentos/engenharia-eletrica/pesquisas-e-projetos/lapsee/downloads/materiais-de-cursos1193/